\documentclass[twocolumn,showpacs]{revtex4-1}
\usepackage{amsfonts}
\usepackage{amssymb}
\usepackage{amsthm}
\usepackage{amsmath}
\usepackage{geometry}
\usepackage{setspace}
\usepackage{mathtools}
\usepackage{enumitem}
\usepackage{mathrsfs}
\usepackage{graphicx}
\usepackage{floatrow}
\usepackage{subcaption}
\usepackage{multirow}
\usepackage{epstopdf}
\usepackage{array}
\newcolumntype{H}{>{\setbox0=\hbox\bgroup}c<{\egroup}@{}}
\usepackage{natbib}
\begin{document}
\title{The Mathematics of Human Contact:\\Developing a Model for Social Interaction in School Children}
\thanks{Based on work presented at the 13th Econophysics Colloquium \& 9th Polish Symposium on Physics in Economy and Social Sciences, Warsaw, July 2017}
\author{Stephen Ashton}
\affiliation{School of Mathematical and Physical Sciences, University of Sussex, Falmer, Brighton, BN1 9QH, United Kingdom}
\author{Enrico Scalas}
\affiliation{School of Mathematical and Physical Sciences, University of Sussex, Falmer, Brighton, BN1 9QH, United Kingdom}
\author{Nicos Georgiou}
\affiliation{School of Mathematical and Physical Sciences, University of Sussex, Falmer, Brighton, BN1 9QH, United Kingdom}
\author{Istv\'{a}n Zolt\'{a}n Kiss}
\affiliation{School of Mathematical and Physical Sciences, University of Sussex, Falmer, Brighton, BN1 9QH, United Kingdom}
\date{\today}

\begin{abstract}
In this paper, we provide a statistical analysis of high-resolution contact pattern data within primary and secondary schools as collected by the SocioPatterns collaboration. Students are graphically represented as nodes in a temporally evolving network, in which links represent proximity or interaction between students.

This article focuses on link- and node-level statistics, such as the on- and off-durations of links as well as the activity potential of nodes and links.

Parametric models are fitted to the on- and off-durations of links, inter-event times and node activity potentials and, based on these, we propose a number of theoretical models that are able to reproduce the collected data within varying levels of accuracy. By doing so, we aim to identify the minimal network-level properties that are needed to closely match the real-world data, with the aim of combining this contact pattern model with epidemic models in future work.
\end{abstract}

\maketitle

\section{Introduction}
The use of networks to model contact patterns or interactions between individuals has proved to be a step change in how epidemics and other spreading processes are modelled \cite{Kiss2017,Latora2017,doi:10.1137/S003614450342480,Widgren2016,Wang2016,Danon2011}. The basic ingredient of such models is to represent individuals by nodes and contacts between these as links between nodes. The use of graph-theoretical methods have helped to reveal and understand the role of contact heterogeneity, preferential mixing and clustering in how disease invade and spread \cite{Bansal879,Keeling295}. Having good network models is crucial. Simple mechanistic models that capture and preserve key properties of empirical networks are often employed as they offer greater flexibility in changing and tuning various network properties. While such models and theory are well developed for static networks, it is only recently that we have empirically measured real-world time varying forms \cite{Rocha2016,Stehle2011,Bansal879,Sah169573,Keeling295,Eubank2005,Enns2011,Steinhaeuser2009,Danon2011,Kiss1332,Skyrms01082000}. 

Current underlying models for network-based epidemiology fall into a handful of categories. Some just use empirical data collected from sensors and apply an appropriate disease model to this \cite{Bansal879,Rocha2016,Stehle2011}. Others use a fairly elementary model where links appear as in gathered data, but are given lifespans drawn from a uniform distribution \cite{Rocha2016}, or are given a simple weighting drawn directly from the data \cite{Sah169573,Stehle2011}. Others take collected data and use it to convert a series of fully connected networks into sparse ones \cite{Sah169573}. Alternative methods involve the use of an idealised network \cite{Keeling295}, regular random network \cite{Bansal879}, random Poisson network \cite{Bansal879,Keeling295}, scale-free random network \cite{Bansal879} or lattice \cite{Keeling295}. In this paper our aim is to analyse an empirical time varying network, in a statistically rigorous way, and build theoretical models that are able to reproduce and mimic the behaviour observed from data.

We will re-analyse data previously collected by the SocioPatterns collaboration (\url{http://www.sociopatterns.org/}) with special focus on time-varying contact patterns in a primary \cite{Gemmetto2014,10.1371/journal.pone.0023176} and high \cite{10.1371/journal.pone.0136497} school. In particular we will focus on measuring properties such as activation time and duration of links as well as off-durations of links. We will then propose and fit candidate parametric distributions to the empirical data. Based on these, we will propose a few different theoretical time-varying network models. Two different model types are proposed. The first model assigns on-off durations to each link from an appropriate probability distribution. Our second model triggers activations at appropriate times (with inter-event times being drawn from an appropriate distribution), before selecting the link to be activated (using a probability matrix drawn from the original data) and assigning an on-duration to that link from an appropriate probability distribution. Even if these models do not capture all the important features of the real-world network, they still provide a useful first approximation. Whilst we focus on school classrooms, our approach can be adapted to modelling other types of social interactions.

\section{Data Collection and Description}
In the original data - both for the primary and high school students - the participants were equipped with sensors that deemed them `in contact' if they were within 1 to 1.5m of each other (an interaction), chosen by the organizers of the original study. This was to act as a proxy of a close-range encounter during which a communicable disease infection can be transmitted, for example, either by cough or sneeze, or directly by hand contact \cite{10.1371/journal.pone.0023176}. Every 20s, a radio packet would be exchanged between the sensors, and all packets transferred would be relayed to a central system to be recorded. This scale was deemed to allow an adequate description of person-to-person interactions that includes brief encounters \cite{10.1371/journal.pone.0023176}.

In both cases, this central system saved the data in a CSV file, with each row containing the timestamp (in 20s intervals), the IDs of the two sensors in contact, and some additional data about the two participants (such as their class). We modify the original data slightly before our initial analysis. Firstly, we remove any participants marked as staff from the data as their behaviour could be potentially anomalous when compared to that of the school children. Whilst we acknowledge that this could remove any potential impact of staff on the behaviour of pupils, we feel justified in this as staff only account for $11$ participants and approximately $5\%$ of the originally recorded links, which may prove problematic in terms of drawing any statistically significant conclusions about potential behaviour. In future work, including this additional layer may lead to an improved model - however, we feel that more data describing these interactions would be needed before we could confidently add this to a model. We also split the students into their separate classes. Whilst this results in the discarding of approximately $20\%$ of the originally recorded links, this given us more samples to analyse; moreover, it allows for a statistical comparison between the dynamics of different classes. From a more practical perspective, this restriction to classes has a considerable impact on the runtime of the model simulations (reducing this size from around 500 students to around 25).

The choice to restrict to classes is also justified from a modelling perspective as it is realistic to assume (at least as an initial hypothesis) that contacts outside of the classroom (during break/lunch) would follow substantially different behaviour.

We also split the data into individual days - similarly to splitting by class, this helped reduce the runtime of the simulation as well as increasing the number of samples we could analyse. Again, this is not unrealistic, as the interactions between students in the same class can reasonably be assumed to be similar from one day to another.

\subsection{Analysis of Original Data}\label{ssec:features}
A series of MATLAB functions were written to take these (separated) CSV files and perform an analysis of a variety of network and temporal features, and attempt to do best-fit analysis on all appropriate results - a full list of these features below. Animations showing the network evolution over time were also produced. For a listing of the code and short descriptions of the functions written to carry out this analysis, please see the handbook provided in the Supplemental Materials.

We identified a variety of key features for analysis. As usual, many more features can be observed from the data, and indeed, in order to approximate a completely realistic model, many of these should be analysed and incorporated into more detailed models. Our models are just an initial step into understanding these socially-interaction temporal networks, and we are only focusing on aspects that categorise and describe both the topology of the network and several temporal properties of the system. These features are presented below, along with brief definitions of these terms:\\
\noindent \textbf{Active Nodes:} The measure of active nodes at a given time $t$ is defined as the number of pupils involved in at least one interaction at time $t$, as a fraction of all pupils active during that day.\\
\textbf{Active Links:} The measure of active links at a given time $t$ is defined as the number of unique (undirected) pupil-pupil interactions at time $t$, as a fraction of all possible links for that day, equal to $\ell_{\mathrm{max}}=N(N-1)/2$, where $N$ is the number of pupils active during that day in the class under consideration.\\
\textbf{Node \& Link Activity Potential:} The activity potential of a node is defined as the number of activations involving that node, as a fraction of all node activations across the day \cite{Perra2012}. We also define an analogue for links, defined as the number of activations of that link, as a fraction of all link activations across the day.\\
\textbf{Global Clustering Coefficient:} The global clustering coefficient at a given time $t$ is defined as the ratio between the number of closed triplets and the number of connected triplets in the network \cite{doi:10.1137/S003614450342480}. That is, the ratio between the number of triangles in the network and the number paths of length $2$, that do not have a third edge connecting the end points.\\
\textbf{Node Degree:} The degree of node $n$ is the number of active links involving it \cite{Diestel2010}.\\
\textbf{Component Features:} Defining a component as a maximal subset of nodes that are fully connected \cite{Kleinberg2010}, we can also examine properties such as component count and nodes and links per component at a given time $t$.\\
\textbf{Activation Time:} For each link, an activation time is measured - defined as the period of time it takes for that specific link to be activated for the first time.\\
\textbf{On-Duration:} For each link, on-durations are measured - defined as the period of time between the activation and deactivation of that link.\\
\textbf{Off-Duration:} For each link, off-durations are measured - defined as the period of time between the deactivation and reactivation of that link.

\subsection{Properties Identified from Original Data}
In the initial part of this article we use the observed data to fit all of the above quantities to certain distributions. These will act as a stepping stone to the second part of this article, in which we develop theoretical models, in an attempt to recreate the observations using Monte Carlo simulations.

As we do not have any explicit theories for the dynamics of any of our chosen properties, we shall test against a series of appropriate common probability distributions \cite[pp.~899--917]{mun2008} and variations on these, representing a range of behaviours defined on the semi-infinite interval $[0,\infty).$ We will be using exponential, gamma, Rayleigh, log-normal, Mittag-Leffler, generalised Pareto and Weibull distibutions. All of these will have best fit parameters chosen using by three different methods - method of moments \cite{Bowman2006}, maximum likelihood estimators \cite{Scholz2006}, and the curve fitting tool in MATLAB (non-linear least squares) - and then compared to the empirical complementary cumulative distribution functions (eCCDFs) of the original data to determine which one is most optimal.

This comparison was achieved by looking at a variety of statistical distances - Kolmogorov-D, Cramer-von-Mises, Kuiper, Watson, Anderson-Darling and modified versions of the Kullback-Leibler and Jensen-Shannon \cite{Stephens1974}. These distances and comparisons were chosen as they emphasise a wide varying range of properties of the distributions to be compared - with, for example, some being more sensitive to changes in the head and tail of the eCCDF, whilst others are more sensitive to changes in the middle. Finding a distribution that had `good' values for all of these distances would indicate that it was a good fit across the entirety of the compared eCCDF.

In over $75\%$ of cases, the curve fitting tool in MATLAB produced the statistically best parameters, with the parameters chosen using this method in the majority of the remaining cases being only slightly different to those produced using a more optimal method. As a result of this, and additionally considering that the method of moments and least likelihood estimation do not work with all of our chosen distributions, we shall conduct any additional analysis using only the curve fitting tool, and only results produced using this method will be presented and used throughout this paper.

\subsubsection{Results from Data}
Below we present a summary of the distributions chosen using the method described above. Best-fit parameters and comparative distances have been excluded for brevity.\\
\noindent \textbf{Active Links:} The optimal tested distribution for the primary school data was Log-Normal, whilst for the high school data, both the Rayleigh and log-normal distributions gave similar fits, with log-normal being slightly more optimal.\\
\textbf{Active Nodes:} The optimal tested distribution for the primary school data was gamma, whilst for the high school data, the gamma and log-normal distributions both gave similar fits, with log-normal being more optimal in all but the most extreme values.\\
\textbf{Node Activity Potential:} In both data sets, gamma and log-normal distributions gave similar fits, with gamma being fractionally better.\\
\textbf{Links per Component:} For the primary school data, gamma and log-normal distributions both gave similar fits, with log-normal being marginally more optimal. Whilst for the high school data, gamma, log-normal and Rayleigh distributions all gave similar fits, with log-normal being slightly better.\\
\textbf{Nodes per Component:} For the primary school data, gamma, log-normal and Rayleigh distributions all gave similar fits, with log-normal being slightly better. Whilst for the high school data, gamma and log-normal distributions both gave similar fits, with no clear optimal distribution.\\
\textbf{Global Clustering Coefficient:} In both data sets, gamma, log-normal and Rayleigh distributions all gave similar fits. For the primary school data there was no clear optimal distribution between these, whilst for the high school data, a gamma distribution was slightly better.\\
\textbf{Interaction Times/On-Times:} In both data sets, the optimal tested distribution was generalised Pareto.\\
\textbf{Number of Components:} In both data sets, the optimal tested distribution was gamma.\\
\textbf{Time Between Contacts/Off Times:} In both data sets, the best tested distribution was log-normal.

\subsubsection{Link Inhomogeneity}
Not surprisingly, the off-durations of links (recalling that links are off if participants ar not in contact with each other) cannot be assumed to be homogeneous across students. This is in accordance with the realistic assumption that certain children are more popular or sociable than others. A further later of statistical fitting determined that for attempting to recreate the primary school data, it was optimal to have the off-durations vary link-by-link. The optimal choice for this was an exponential distribution with log-normal parameters. We additionally examined the \textbf{triangle count} within the network, as well as \textbf{inter-event times} (the time between two consecutive link activations in the network). For the first, a gamma distribution was the optimal fitted distribution, whilst for the second, a log-normal distribution was selected.

These two features were chosen to be added to the list of those analysed as the triangle count offers an additional measurement of the nature of the network structure alongside the global clustering coefficient, whilst the inter-event times were necessary for building our second model. 

\subsubsection{Comparing Samples}\label{sssec:comparesamples}
\begin{table}
\centering
\begin{tabular}{l|c|c|H} \cline{2-3}
& \textbf{1\%} & \textbf{5\%} & \textbf{10\%}\\ \cline{1-3}
\multicolumn{1}{|l|}{\textbf{Active Links}} & $10$ & $3$ & $3$\\ \cline{1-3}
\multicolumn{1}{|l|}{\textbf{Active Nodes}} & $139$ & $125$ & $112$\\ \cline{1-3}
\multicolumn{1}{|l|}{\textbf{Node Activity Potential}} & $190$ & $185$ & $179$\\ \cline{1-3}
\multicolumn{1}{|l|}{\textbf{Global Clustering Coefficient}} & $67$ & $46$ & $38$\\ \cline{1-3}
\multicolumn{1}{|l|}{\textbf{Interaction Time}} & $9$ & $4$ & $3$\\ \cline{1-3}
\multicolumn{1}{|l|}{\textbf{Time Between Contacts}} & $30$ & $19$ & $14$\\ \cline{1-3}
\multicolumn{1}{|l|}{\textbf{Component Count}} & $113$ & $94$ & $87$\\ \cline{1-3}
\multicolumn{1}{|l|}{\textbf{Links per Component}} & $62$ & $50$ & $45$\\ \cline{1-3}
\multicolumn{1}{|l|}{\textbf{Nodes per Component}} & $63$ & $54$ & $47$\\ \cline{1-3}
\multicolumn{1}{|l|}{\textbf{Triangle Count}} & $114$ & $99$ & $95$\\ \cline{1-3}
\end{tabular}

\caption{Acceptances of $\mathcal{H}_{0}$ (see equation \ref{eq:h0comparesamples}) at $1\%$ and $5\%$ Levels (max: 190) (see subsection \ref{sssec:comparesamples} for full explanation)}
\label{tab:comparesamples}
\end{table}
When we create our models, we aim to have little dependence on the original data - varying parameters only between differing settings (primary school vs. high school), rather than within these settings. For example, we would aim to have the parameters for the random variable generation for the model for class 5A in the primary school to be the same as those in the model for class 1B of the primary school. Therefore, our first statistical test will be to test the validity of this statement. Our $\mathcal{H}_{0}$ is
\begin{equation}\label{eq:h0comparesamples}
\begin{split}
\mathcal{H}_{0}:  &\text{ The two observed samples come}\\
									&\text{ from a common distribution.}
\end{split}
\end{equation}
We compute two-sample Kolmogorov-Smirnov distances \cite{smirnov1939estimate,Press1992} between each of our  data sets within each setting.

We present the number of acceptances of this hypothesis (out of 190) for our primary school data samples at the 1 and 5 percent levels in Table \ref{tab:comparesamples}. 
Examining these results, we conclude that while we do not have a unanimous degree of acceptances for $\mathcal{H}_{0},$ we have a substantial number in some metrics and a notable level in others. Other metrics have a very low degree of matching - most noticeably in terms of active links and interaction times. Whilst this is not ideal for our aim to only vary parameters between scenarios, for brevity we shall still proceed under this assumption - although it should be noted that when we present our models we do not actually fix the parameter in the distribution for our interaction times. Instead we draw this parameter from a random distribution itself, which reflects this behaviour in the data originally collected by the SocioPatterns Collaboration.

\section{Model Creation}\label{sec:models}
The aim of our model is to recreate the dynamics seen in the original data with as few properties and parameters taken from the original data as possible. In more precise terms, we wish to test if the mechanism of interactions within the original data can be explained by a small number of key factors and identify and refine those parameters. As with any model, we doubt that we will be able to replicate every property in the original data, but it is important to examine the differences between our model and the original data, and to put a measurement on the distance between the two. Whilst there will be some properties that we will be controlling, there will be several network and temporal properties that emerge from our model that we can compare to our original data, hence giving us a measure of the distance between the two. For the sake of brevity, we will only present the results and parameter values for primary school data below. Analysis supporting our choice of distributions and parameters is provided in the Supplemental Materials.

\subsection{Model 1}
For this stage-0 model we look at each (potential) link individually and model its behaviour as an alternating renewal process (ARP). We also include an initialization phase for each link that models the time (in seconds) until the first activation of that link. This can be seen as the following process for each link where $X_{ij,n}$ represents duration of the $n$-th on (or off) phase for the link $(i, j)$, with the distributions chosen using an empirical analysis of the data. Algorithmically, we present this as:
\begin{enumerate}
\item \textbf{Initialization Phase:} Generate the initialisation time for this link with $$X_{ij}^{\mathrm{Init}} \sim \text{Exp}(6278.0)$$
\item \textbf{ARP On-Phase:} Assign the link the on-duration $$X_{ij,n}^{\mathrm{On}} \sim \text{Exp}\left(Y_{ij}\right)$$ with parameter fixed for each $(i, j)$ to $$Y_{ij} \sim \text{LogNormal}(3.5348,0.2807).$$
\item \textbf{ARP Off-Phase:} Assign the link the off-duration as $$X_{ij,n}^{\mathrm{Off}} \sim \text{LogNormal}(6.3512,1.3688).$$
\item \textbf{Repeating Process:} Repeat Stages 2 and 3 until the total time has reached or exceeded the simulation time.
\end{enumerate}
\subsection{Model 2a}
In this stage-0 model, we will be dealing with the system on a macroscopic basis. We are drawing times between activations from an appropriate distribution, then at each of these activations, a link is chosen at random from a custom distribution constructed from the link activity potentials (as defined in subsection \ref{ssec:features}) extracted from the data and represented by a symmetric weighting matrix $M$. If the chosen link is already active in the network, this selection is discarded, and another link is chosen for that activation time. Once a link has been activated, it is given a lifespan from an appropriate distribution. This can be seen as the following process, with the distributions chosen using an empirical analysis of the data. Algorithmically, we present this as:
\begin{enumerate}
	\item \textbf{Time between Activations:} Generate $$t_{i} \sim \mathrm{LogNormal}(5.6901\times 10^{-4},1.7957).$$
	\item \textbf{Link Activation:} At each activation time $T_{k}$, defined as $$T_{k}=\sum_{i=0}^{k}{t_{i}},$$ a link $(n_{1},n_{2})$ is chosen using the relative weights in the matrix $M$. If $(n_{1},n_{2})$ is already active at time $T_{k}$, choose another link for this time $(n'_{1},n'_{2})$.
	\item \textbf{Assign On-Durations:} This link is given the duration $$X_{n_1 n_2}^{k} \sim \mathrm{Exp}(Y_{n_1 n_2})$$ as before with parameter fixed for each $(n_1,n_2)$ to  $$Y_{n_1 n_2} \sim \mathrm{LogNormal}(3.5348,0.2807).$$
\end{enumerate}
\subsection{Model 2b}\label{ssec:model2bcreation}
In this model, we modify our Model 2a and attempt to improve triangle count and clustering. Most of the method is similar to the earlier model, but we force chosen links to close a pair of links into a triangle at a fixed rate, reweighting our selection matrix to only account for these links (if no such links exist, we use the original selection matrix), before proceeding as before with this link selected. This can be seen as the following algorithm, with the distributions always chosen using an empirical analysis of the data:
\begin{enumerate}
	\item \textbf{Time between Activations:} Generate $$t_{i} \sim \mathrm{LogNormal}(5.6901\times 10^{-04},1.7957).$$
	\item \textbf{Triangulation Bias:} Generate a random number $u$ such that
	$$u \sim \mathrm{Unif}[0,1].$$
	If $u\geq 0.0640$ (our `forcing' rate, calculated from the data), proceed to Stage 3a, else proceed to Stage 3b.
	\item \textbf{Link Activation:}
	\begin{enumerate}
	\item \textbf{Standard Activation:} At each activation time $T_{k}$, defined as $$T_{k}=\sum_{i=0}^{k}{t_{i}},$$ a link $(n_{1},n_{2})$ is chosen using the relative weights in the matrix $M$. If $(n_{1},n_{2})$ is already active at time $T_{k}$, choose another link for this time $(n'_{1},n'_{2})$. Proceed to Stage 4.
	\item \textbf{Triangle-Biased Activation:}
		\begin{enumerate}
		\item \textbf{Matrix Reweighting:} Generate the (symmetric logical) matrix $C$ of links that will complete triangles. If this matrix is $0$, set $C=\mathbb{I}$. Create the adjusted weighted matrix $M'$ where $M'_{ij}=C_{ij}M_{ij}$.
		\item \textbf{Link Activation:} At each activation time $T_{k}$, defined as $$T_{k}=\sum_{i=0}^{k}{t_{i}},$$ a link $(n_{1},n_{2})$ is chosen using the relative weights in the adjusted matrix $M'$. If $(n_{1},n_{2})$ is already active at time $T_{k}$, choose another link for this time $(n'_{1},n'_{2})$. Proceed to Stage 4.
	\end{enumerate}
	\end{enumerate}
	\item \textbf{Assign On-Durations:} This link is given the duration $$X_{n_1 n_2}^{k} \sim \mathrm{Exp}(Y_{n_1 n_2})$$ as usual with parameter fixed for each $(n_1,n_2)$ to $$Y_{n_1 n_2} \sim \mathrm{LogNormal}(3.5348,0.2807).$$
\end{enumerate}
\subsection{Model 2c}
We shall again build upon our previous model - Model 2b - this time changing our matrix $M$. Previously, this has been a fixed matrix extracted from the data, but we wish to move to a randomly generated one to reduce this strict dependency on the original data. Analysing these (symmetric) matrices, we examine the row (or column) sums, which we attempt to find a distribution for. From an analysis of the data, we choose an appropriate distribution for these sums - we shall use row sums $$M_{i\Sigma} = \sum_{j=1}^{n}{M_{ij}} \sim \Gamma(12.3109,0.0037).$$
For our first attempt at generating an appropriate random matrix $M$, we shall assume that each term is taken from a gamma distribution with
$$M_{ij}\sim\Gamma(\mu^{A}_{i},0.0037)+\Gamma(\mu^{B}_{j},0.0037)$$
for $i<j$, $M_{ij}=0$ for $i=j$ and $M_{ij}=M_{ji}$ for $i>j$. This distribution is chosen in a simple yet natural way that ensures correlations across rows and columns. We also construct this in such a way that the choice of a self-loop is impossible, whilst also ensuring symmetry (which is to be expected as our network is undirected). Due to the additive properties of the gamma distribution, this is equivalent to the distribution
$$M_{ij}\sim\Gamma(\mu^{A}_{i}+\mu^{B}_{j},0.0037)$$
for $i<j$, $M_{ij}=0$ for $i=j$ and $M_{ij}=M_{ji}$ for $i>j$.

We can use the properties of the gamma distribution to specify the parameters $\mu^A_i$ and $\mu^B_j$ as follows. As this matrix has to be symmetric, we modify those entries below the diagonal accordingly.
To sum across a row, we first add the entries to the right of the diagonal, which is equal to
$$(n-i)\mu^{A}_{i} + \sum_{j=i+1}^{n}\mu^{B}_{j}.$$
We then notice that the entries to the left of the diagonal, are equal to the column sum to the diagonal, equal to
$$(i-1)\mu^{B}_{i} + \sum_{j=1}^{i-1}\mu^{A}_{j},$$
giving to total sum to be
$$(n-i)\mu^{A}_{i} + \sum_{j=i+1}^{n}\mu^{B}_{j} + (i-1)\mu^{B}_{i} + \sum_{j=1}^{i-1}\mu^{A}_{j}.$$
To match the distributions for the row sums, we require that:
\begin{align*}
(n&-1)\mu^{A}_{1} + \sum_{j=2}^{n}\mu^{B}_{j}\\
&= (n-2)\mu^{A}_{2} + \sum_{j=3}^{n}\mu^{B}_{j} + \mu^{B}_{2} + \mu^{A}_{1}\\
&= (n-3)\mu^{A}_{3} + \sum_{j=4}^{n}\mu^{B}_{j} + 2\mu^{B}_{3} + \sum_{j=1}^{2}\mu^{A}_{j}\\
&= \hdots \\
&= \mu^{A}_{n-1} + \mu^{B}_{n} + (n-2)\mu^{B}_{n-1} + \sum_{j=1}^{n-2}\mu^{A}_{j}\\
&= (n-1)\mu^{B}_{n} + \sum_{j=1}^{n-1}\mu^{A}_{j} = 12.3109
\end{align*}
The trivial solution to this is $\mu^{A}_{i}=\mu^{B}_{j}=\mu^{\star} \; \forall i,j\in\{1,2,\hdots,n\}$, giving $\mu^{\star}=12.3109/2(n-1)$. Our initial model for a randomly generated symmetric $M$ shall be with $$M_{ij} \sim \Gamma\left(\frac{12.3109}{2(n-1)},0.0037\right)$$ for $i<j$, $M_{ij}=0$ for $i=j$ and $M_{ij}=M_{ji}$ for $i>j$. Whilst the use of this trivial solution is somewhat simplistic, we believe that the inclusion of this method is an important step as it allows us to examine behaviours and test mechanics before examining non-trivial solutions in future work.

\subsection{Summary}\label{ssec:summary}
\begin{table*}
\centering
\begin{tabular}{|c|c|c|c|}\hline
\textbf{Model} & \textbf{Parameters} & \textbf{Parameter Values} & \textbf{Parameter Count} \\ \hline
\multirow{3}{*}{Model 1}	&		$X_{ij}^{\mathrm{Init}} \sim \text{Exp}(\lambda)$	& $\lambda = 6278.0$																	& \multirow{3}{*}{5} 	\\ \cline{2-3}
													&		$Y_{ij} \sim \text{LogNormal}(\mu_{1},\sigma_{1}^{2})$ & $(\mu_{1},\sigma_{1}^{2}) = (3.5348,0.2807)$ 		&											\\ \cline{2-3}
													&		$X_{ij,n}^{\mathrm{Off}} \sim \text{LogNormal}(\mu_{2},\sigma_{2}^{2})$ & $(\mu_{2},\sigma_{2}^{2}) = (6.3512,1.3688)$ 		&	\\ \hline\hline
\multirow{2}{*}{Model 2a}	&		$t_{i} \sim \mathrm{LogNormal}(\mu_{1},\sigma_{1}^{2})$	& $(\mu_{1},\sigma_{1}^{2}) = (5.6901\times 10^{-4},1.7957)$	& \multirow{2}{*}{4} 	\\ \cline{2-3}
													&		$Y_{n_1 n_2} \sim \mathrm{LogNormal}(\mu_{2},\sigma_{2}^{2})$	& $(\mu_{2},\sigma_{2}^{2}) = (3.5348,0.2807)$			& 	\\ \hline \hline
\multirow{4}{*}{Model 2b}	&		$t_{i} \sim \mathrm{LogNormal}(\mu_{1},\sigma_{1}^{2})$	& $(\mu_{1},\sigma_{1}^{2}) = (5.6901\times 10^{-4},1.7957)$	& \multirow{4}{*}{$5+\frac{n(n-1)}{2}$} 	\\ \cline{2-3}
													&		$Y_{n_1 n_2} \sim \mathrm{LogNormal}(\mu_{2},\sigma_{2}^{2})$	& $(\mu_{2},\sigma_{2}^{2}) = (3.5348,0.2807)$			& 	\\ \cline{2-3}
													&		$u \geq u_{f}$ (our `forcing' rate) & $u_{f}=0.0640$ & \\ \cline{2-3}
													&		$M$ & $n\times n$ symmetric matrix & \\ \hline \hline
\multirow{4}{*}{Model 2c}	&		$t_{i} \sim \mathrm{LogNormal}(\mu_{1},\sigma_{1}^{2})$	& $(\mu_{1},\sigma_{1}^{2}) = (5.6901\times 10^{-4},1.7957)$	& \multirow{4}{*}{7} 	\\ \cline{2-3}
													&		$Y_{n_1 n_2} \sim \mathrm{LogNormal}(\mu_{2},\sigma_{2}^{2})$	& $(\mu_{2},\sigma_{2}^{2}) = (3.5348,0.2807)$			& 	\\ \cline{2-3}
													&		$u \geq u_{f}$ (our `forcing' rate) & $u_{f}=0.0640$ & \\ \cline{2-3}
													&		$M_{ij} \sim \Gamma(k,\theta)$ & $(k,\theta)=\left(\frac{12.3109}{2(n-1)},0.0037\right)$ & \\ \hline
\end{tabular}
\caption{Summary of Model Dependencies (see subsection \ref{ssec:summary} for full explanation and subsection \ref{ssec:model2bcreation} for the definitions of $u_{f}$ and $M$)}
\label{tab:summary}
\end{table*}
In Table \ref{tab:summary} we present a concise comparative summary of the data dependencies of each of our 4 model variants. For most of our models, we feel as though the parameter count is acceptable considering the complexities of the behaviours we are attempting to capture. In Model 2b, the parameter count is much higher than reasonable due to the explicit dependence on the original data, suggesting that this would not be an ideal model to fully implement - however it is included in our analysis in order to allow us to observe the accuracy of Model 2c.

\section{Model Analysis}\label{sec:modelanalysis}
\begin{table*}
\centering
\begin{tabular}{ccHH|c|c|HHc|c|c|c|c|} \cline{5-12}
& & \rotatebox[origin=c]{90}{\parbox{2.2cm}{\textbf{Active Links}}} & \rotatebox[origin=c]{90}{\parbox{2.2cm}{\textbf{Active Nodes}}} & \rotatebox[origin=c]{90}{\textbf{\parbox{2.2cm}{Node Activity Potential}}} & \rotatebox[origin=c]{90}{\parbox{2.2cm}{\textbf{Global Clustering Coefficient}}} & \rotatebox[origin=c]{90}{\parbox{2.2cm}{\textbf{Interaction Time}}} & \rotatebox[origin=c]{90}{\parbox{2.2cm}{\textbf{Time Between Contacts}}} & \rotatebox[origin=c]{90}{\parbox{2.2cm}{\textbf{Component Count}}} & \rotatebox[origin=c]{90}{\parbox{2.2cm}{\textbf{Links per Component}}} & \rotatebox[origin=c]{90}{\parbox{2.2cm}{\textbf{Nodes per Component}}} & \rotatebox[origin=c]{90}{\parbox{2.2cm}{\textbf{Triangle Count}}} \\ \hline
\multicolumn{1}{|c|}{\multirow{4}{*}{\textbf{Model 1}}} & \multicolumn{1}{c}{min} & $0.07058$ & $0.05909$ & $0.1304$ & $0.02517$ & $0.1045$ & $0.08932$ & $0.01813$ & $0.009406$ & $0.006316$ & $0.02394$\\ \cline{2-12}
\multicolumn{1}{|c|}{} & \multicolumn{1}{c}{max} & $0.6027$ & $0.6479$ & $0.5769$ & $0.2876$ & $0.2723$ & $0.2996$ & $0.6083$ & $0.07935$ & $0.07935$ & $0.2876$\\ \cline{2-12}
\multicolumn{1}{|c|}{} & \multicolumn{1}{c}{mean} & $0.333$ & $0.3398$ & $0.3365$ & $0.1191$ & $0.1844$ & $0.1889$ & $0.2895$ & $0.03934$ & $0.03856$ & $0.1189$\\ \cline{2-12}
\multicolumn{1}{|c|}{} & \multicolumn{1}{c}{mode} & $0.07058$ & $0.05909$ & $0.3043$ & $0.03524$ & $0.1045$ & $0.08932$ & $0.01813$ & $0.009406$ & $0.006316$ & $0.03524$\\ \hline \hline
\multicolumn{1}{|c|}{\multirow{4}{*}{\textbf{Model 2a}}} & \multicolumn{1}{c}{min} & $0.06973$ & $0.1087$ & $0.08696$ & $0.01835$ & $0.1086$ & $0.05501$ & $0.03448$ & $0.004412$ & $0.00467$ & $0.002869$\\ \cline{2-12}
\multicolumn{1}{|c|}{} & \multicolumn{1}{c}{max} & $0.6398$ & $0.6559$ & $0.4249$ & $0.2677$ & $0.2893$ & $0.2255$ & $0.6067$ & $0.09015$ & $0.09015$ & $0.2641$\\ \cline{2-12}
\multicolumn{1}{|c|}{} & \multicolumn{1}{c}{mean} & $0.4142$ & $0.4265$ & $0.2212$ & $0.09876$ & $0.1909$ & $0.1099$ & $0.308$ & $0.04169$ & $0.04195$ & $0.06908$\\ \cline{2-12}
\multicolumn{1}{|c|}{} & \multicolumn{1}{c}{mode} & $0.06973$ & $0.1087$ & $0.1739$ & $0.01835$ & $0.1086$ & $0.05501$ & $0.03448$ & $0.004412$ & $0.00467$ & $0.002869$\\ \hline \hline
\multicolumn{1}{|c|}{\multirow{4}{*}{\textbf{Model 2b}}} & \multicolumn{1}{c}{min} & $0.07336$ & $0.07336$ & $0.08$ & $0.02513$ & $0.1108$ & $0.05846$ & $0.03112$ & $0.004466$ & $0.003064$ & $0.005029$\\ \cline{2-12}
\multicolumn{1}{|c|}{} & \multicolumn{1}{c}{max} & $0.6555$ & $0.6484$ & $0.4377$ & $0.2065$ & $0.3043$ & $0.24$ & $0.5274$ & $0.08203$ & $0.08203$ & $0.2011$\\ \cline{2-12}
\multicolumn{1}{|c|}{} & \multicolumn{1}{c}{mean} & $0.3986$ & $0.4038$ & $0.2241$ & $0.08137$ & $0.1917$ & $0.118$ & $0.2838$ & $0.0373$ & $0.03754$ & $0.05605$\\ \cline{2-12}
\multicolumn{1}{|c|}{} & \multicolumn{1}{c}{mode} & $0.07336$ & $0.07336$ & $0.2273$ & $0.06793$ & $0.1108$ & $0.05846$ & $0.03112$ & $0.004466$ & $0.003064$ & $0.005029$\\ \hline \hline
\multicolumn{1}{|c|}{\multirow{4}{*}{\textbf{Model 2c}}} & \multicolumn{1}{c}{min} & $0.09512$ & $0.1199$ & $0.08696$ & $0.03049$ & $0.1186$ & $0.06514$ & $0.03836$ & $0.005304$ & $0.006754$ & $0.003863$\\ \cline{2-12}
\multicolumn{1}{|c|}{} & \multicolumn{1}{c}{max} & $0.6688$ & $0.6728$ & $0.4945$ & $0.2036$ & $0.2977$ & $0.2309$ & $0.552$ & $0.08004$ & $0.08004$ & $0.1842$\\ \cline{2-12}
\multicolumn{1}{|c|}{} & \multicolumn{1}{c}{mean} & $0.4161$ & $0.4267$ & $0.2485$ & $0.07893$ & $0.1989$ & $0.1233$ & $0.3087$ & $0.04149$ & $0.04166$ & $0.04898$\\ \cline{2-12}
\multicolumn{1}{|c|}{} & \multicolumn{1}{c}{mode} & $0.5195$ & $0.1199$ & $0.2273$ & $0.07942$ & $0.1186$ & $0.06514$ & $0.03836$ & $0.005304$ & $0.006754$ & $0.006151$\\ \hline
\end{tabular}

\caption{Selected Two-Sample Kolmogorov-Smirnov Distances (see section \ref{sec:modelanalysis} for full explanation)}
\label{tab:distances}
\end{table*}
Please note, in the figures highlighting key results, simulated data is represented by crosses whereas observed data is represented by dotted lines, with the data displayed as an eCCDF with log-log axes (with scaling preserved between models). Each colour represents a different simulation or data set. In order, the four eCCDFs shown represent active nodes, node activity potentials, component counts and the global clustering coefficients. We choose these metrics to illustrate as they represent both promising behaviours and less-optimal ones, thereby giving a representative snapshot of our results. Additionally, these eCCDFs are some of the clearer and easier ones to read, allowing us to demonstrate a number of behaviours in a brief and compact manner. It should be noted that in some cases (most apparent in the case of the global clustering coefficients) that some of these eCCDFs appear not to start at $1$ as expected - this is a result of a high prevalence of the value $0$ in our data, with a large jump between this and other values. For readability, this jump has been excluded from the graphics, with our images only showing the section of the graph where the majority of our values fall.

We also present comparative data in two tables. In Table \ref{tab:distances}, we show a summary of the two-sample Kolmogorov-Smirnov distances \cite{smirnov1939estimate,Press1992} between our collection of 20 empirical samples and 20 simulated data samples from each of the 4 models presented above - showing the minimum, maximum, mean and mode of the distance between any of the 20 sets of real world data and any of the 20 sets of generated data.
We also compare horizontally, comparing each empirical data set against 50 data sets generated using our chosen metrics. We test the hypothesis $\mathcal{H}_{0},$ in this case, this is
\begin{equation}\label{eq:h0comparisons}
\begin{split}
\mathcal{H}_{0}:  &\text{ The chosen empirical and generated}\\
									&\text{ data samples come from a common}\\
									&\text{ distribution.}
\end{split}
\end{equation}
In Table \ref{tab:comparisons}, we present the total number of acceptances (out of a possible 1000) at the $5\%$-level of this hypothesis when tested on a particular metric.
\begin{center}\begin{figure}
\begin{subfigure}{0.4\columnwidth}
\includegraphics[width=\columnwidth]{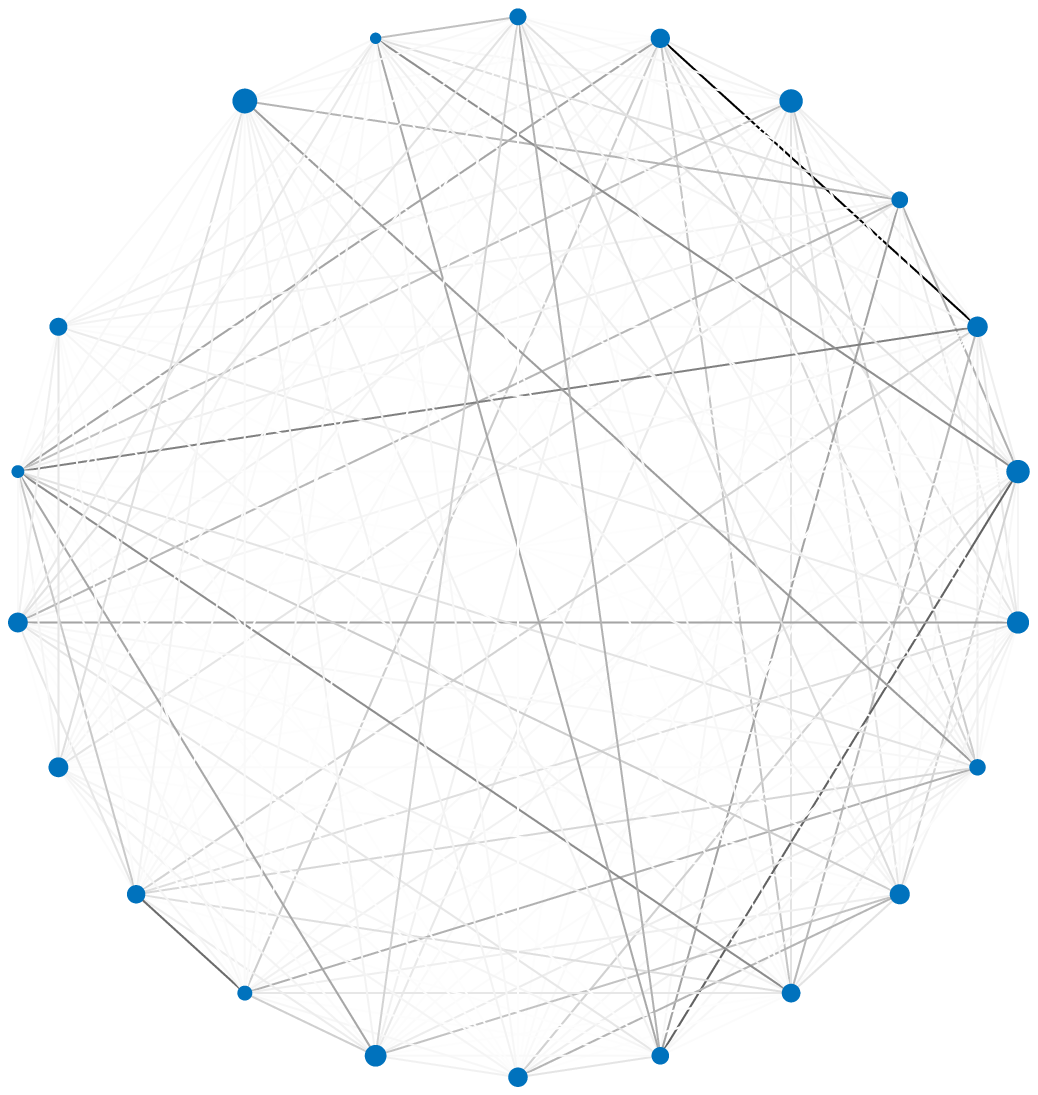}
        \caption{Original Data}
        \label{subfig:echoes_orig}
\end{subfigure}
\begin{subfigure}{0.4\columnwidth}
\includegraphics[width=\columnwidth]{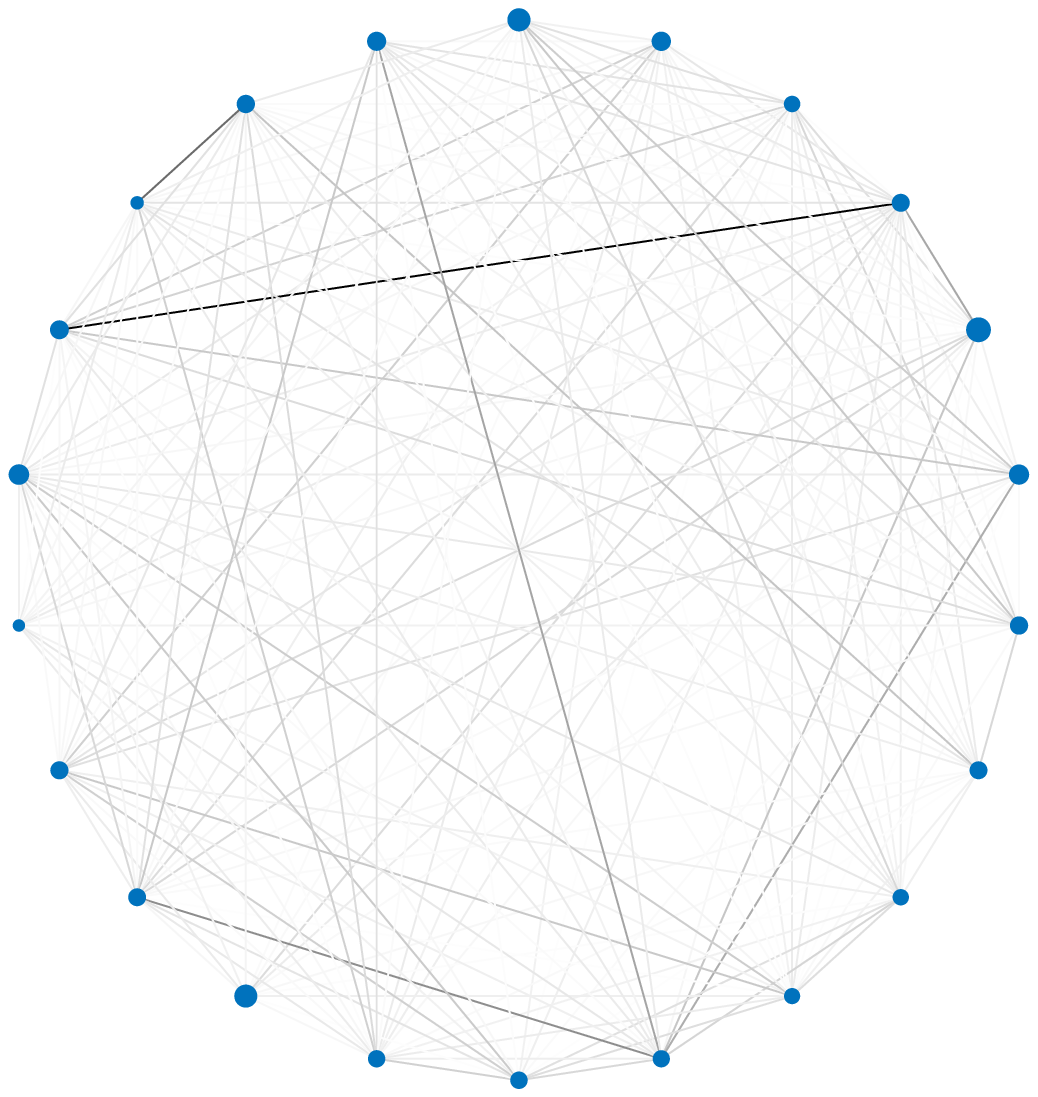}
        \caption{Model 1}
        \label{subfig:echoes_m1}
\end{subfigure}

\begin{subfigure}{0.4\columnwidth}
\includegraphics[width=\columnwidth]{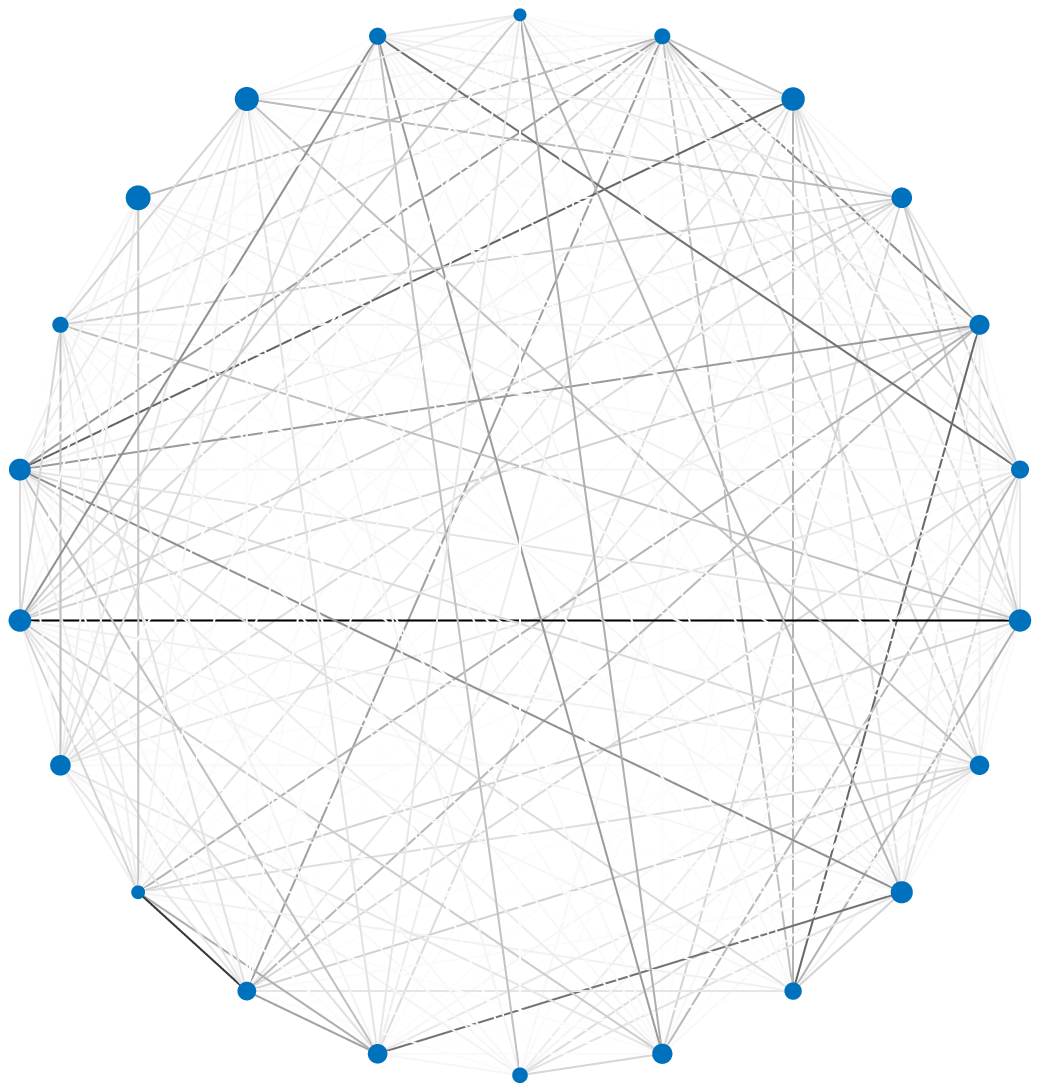}
        \caption{Model 2a}
        \label{subfig:echoes_m2a}
\end{subfigure}
\begin{subfigure}{0.4\columnwidth}
\includegraphics[width=\columnwidth]{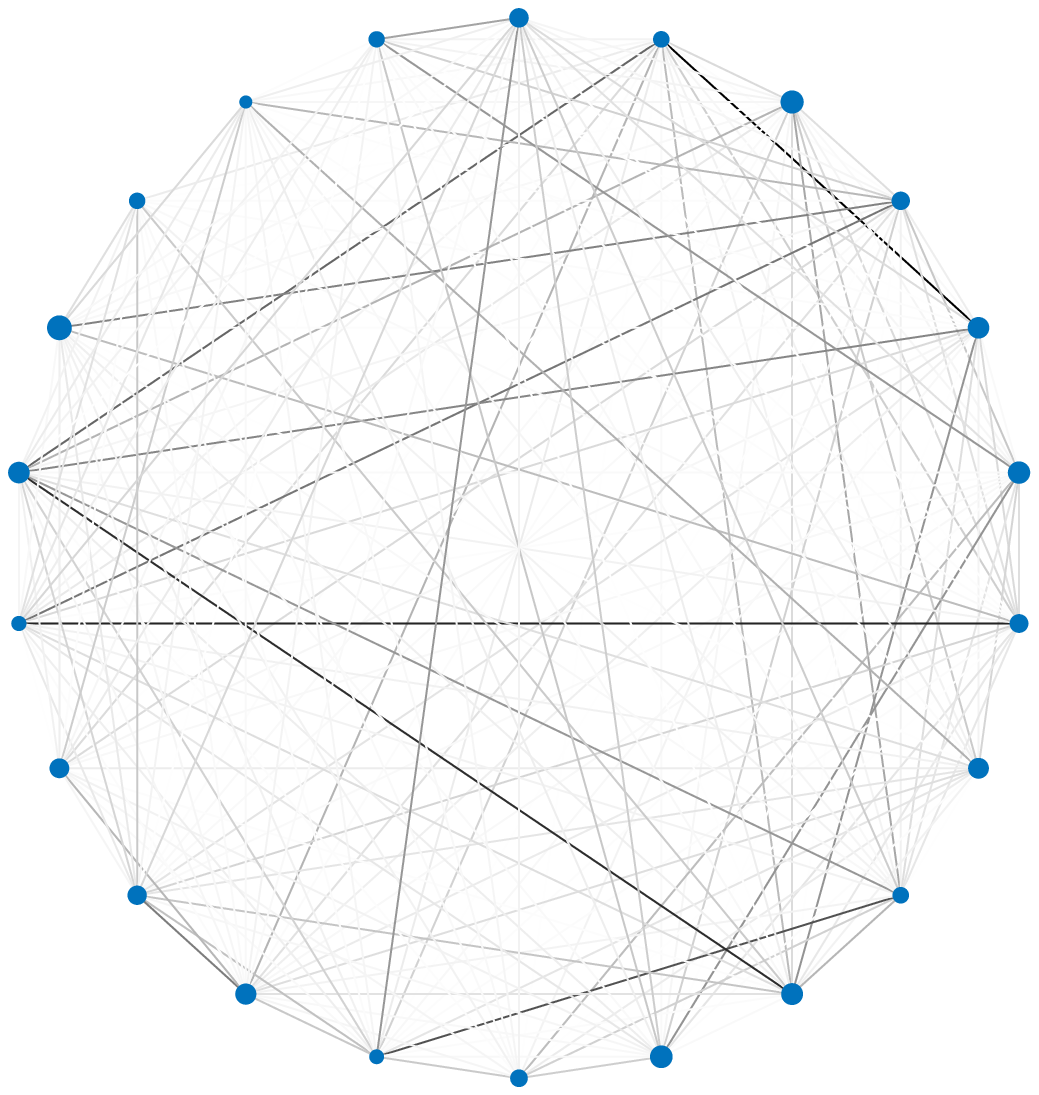}
        \caption{Model 2b}
        \label{subfig:echoes_m2b}
\end{subfigure}

\begin{subfigure}{0.4\columnwidth}
\includegraphics[width=\columnwidth]{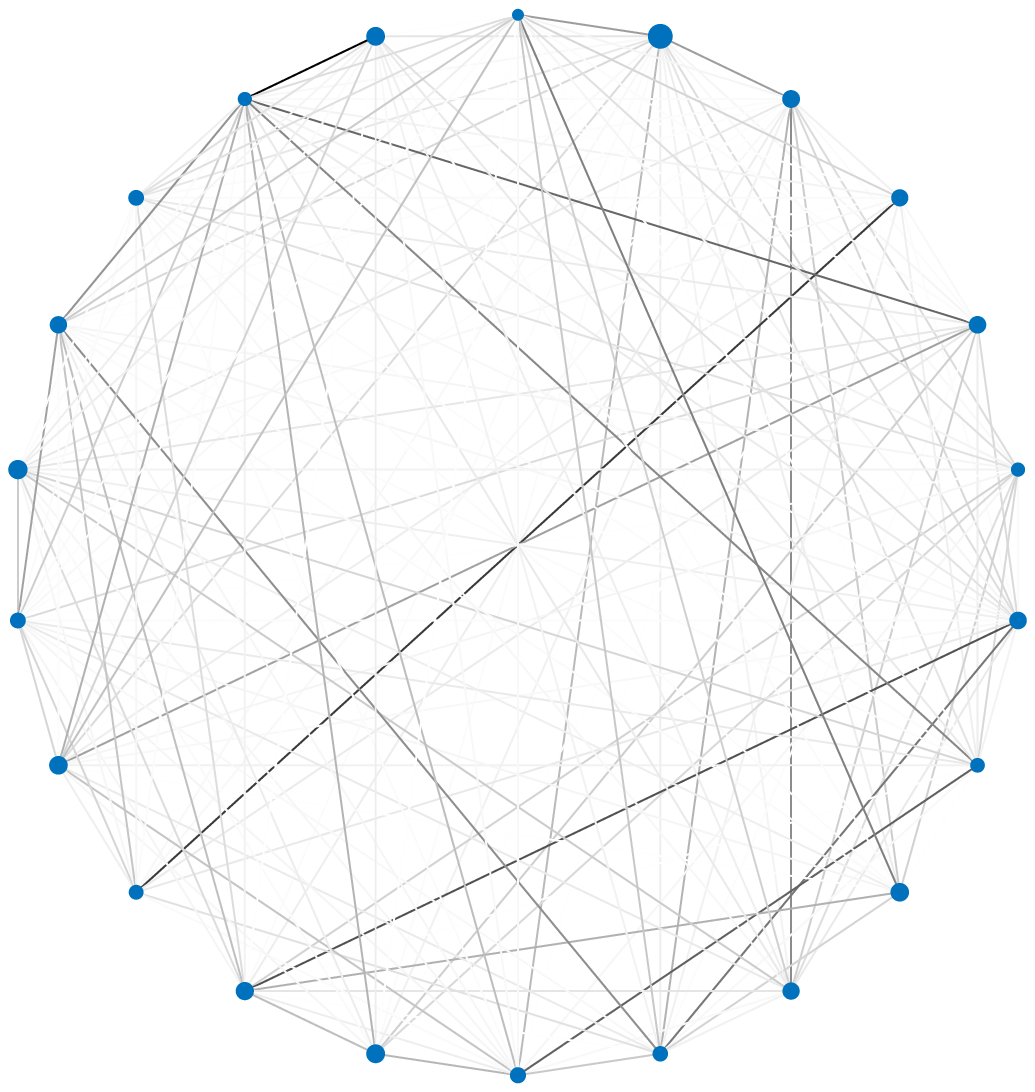}
        \caption{Model 2c}
        \label{subfig:echoes_m2c}
\end{subfigure}
\caption{Long Term Behaviours for Original Data and Models - Size of Nodes \& Transparency of Links Represent Relative Activities (see last paragraph of the opening of section \ref{sec:modelanalysis} for full explanation and the relevant subsections of section \ref{sec:modelanalysis} and section \ref{sec:modelcomp} for an analysis of these results). One immediate observation is that Model 1 homogenises much faster - note the limited number of darker links.}
\label{fig:echoes}
\end{figure}\end{center}

Additionally, we present Figure \ref{fig:echoes} to highlight long-term behaviours in our model. In this figure, the transparency of each link represents its relative activity in comparison to other links, and the size of each node represents the relative activity of each node. The 5 images in this figure represent this behaviour at $t=15000$ seconds for an example of the original data, Model 1, Model 2a, Model 2b and Model 2c. Using this figure, we can see systemic behaviours, such as possible grouping of nodes into friendship groups or similar metrics that would be more difficult to measure empirically. This also gives us an intrinsic definition for link \textbf{spread}. Figure \ref{subfig:echoes_m1} demonstrates a poor spread - the long-term behaviour is relatively homogeneous with fewer darker links. Similarly, a simulation that resulted in long-term behaviour that only had darker links limited to a very small number of nodes would also suffer from poor spread. Comparatively, Figure \ref{subfig:echoes_orig} has a better edge spread - there are a higher number of darker links spread among a larger number of nodes. More precisely, this is measuring a combination of factors - including activity potentials, component structures and other network features - but allows us to get an impression of many of these features at a glance. We do not expect a perfect matching between the examples here due to the randomness of the data, but are instead looking for system-wide similarity in behaviour. Differences are expected in the placement of stronger links and nodes (and indeed, do occur between simulation runs). However, we would expect a well-fitting model to exhibit similar numbers to those in the original data and with a similar relationship between them (for example, as Figure \ref{subfig:echoes_orig} has many nodes being involved in at least one stronger link, a well-fitting model would not be expected to have all of its strong links emanating from a common node).

\subsection{Model 1}\label{ssec:model1}
\begin{figure}
\centering
\includegraphics[width=0.85\columnwidth]{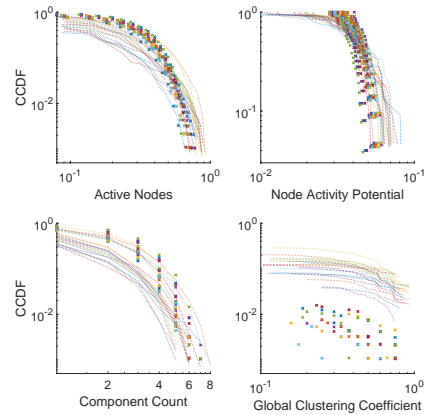}
\caption{Selected Results for Model 1. Simulated data is represented by crosses whereas observed data is represented by dotted lines. Each colour represents a different simulation or data set. See subsection \ref{ssec:model1} for full explanation.}
\label{fig:model1}
\end{figure}
Looking at Figure \ref{fig:model1}, the appropriate sections of Tables \ref{tab:distances} and \ref{tab:comparisons} and other comparative and graphical results not directly presented in this paper for brevity, as a first attempt at creating a model, we see promising results. The model produces acceptable fits for several of the examined features. Active links, active nodes and on-durations all produce graphically acceptable results, although using our Kolmogorov-Smirnov acceptances (shown in Table \ref{tab:comparisons}), there are improvements to be made in terms of these fits. For off-durations, when we compare our eCCDFs, we observe a reasonable fit in certain areas of the distibution although this fit deteriorates for extreme values and once again we notice that our acceptances indicate that the current construction of this model requires refinement to fully capture this behaviour. For our global clustering coefficient (presented in Figure \ref{fig:model1}) and triangle count, we have poor fits where comparing the data sets graphically, although we are getting a small number of acceptances with our two-sample Kolmogorov-Smirnov tests - likely as a result of an extreme prevalence of certain values in these data sets.
For nodes per component, links per component and the component count (partially presented in Figure \ref{fig:model1}), we observe acceptable fits graphically and are indeed accepting a small number of these fits when calculating our statistical distances, as shown in Table \ref{tab:comparisons}. This also indicated indicates that slight refinement to this fit may be possible. For node activity potential, we have a good fit, both graphically and when considering our number of Kolmogorov-Smirnov acceptances.

It is evident that this model does have noticeable differences to the observed data. We have a substantial number of small linear components in our model, which is impacting many of the features described above. Additionally there are problems with link selection spread (defined in section {sec:modelanalysis}) as can be seen when comparing the original behaviour displayed in Figure \ref{subfig:echoes_orig} with that in Figure \ref{subfig:echoes_m1}, resulting in very few popular links (reflecting strong friendships), which could also explain differences within the node activity potentials at the tail of our CCDFs.

\subsection{Model 2a}\label{ssec:model2a}
\begin{figure}
\centering
\includegraphics[width=0.85\columnwidth]{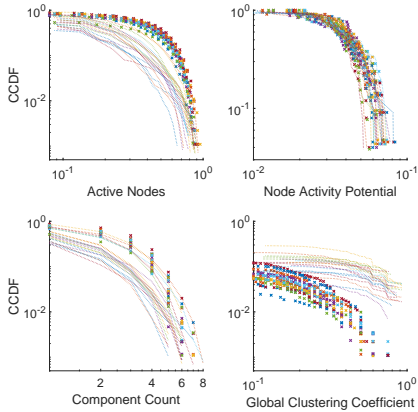}
\caption{Selected Results for Model 2a. Simulated data is represented by crosses whereas observed data is represented by dotted lines. Each colour represents a different simulation or data set. See subsection \ref{ssec:model2a} for full explanation.}
\label{fig:model2a}
\end{figure}
Considering Figure \ref{fig:model2a}, the appropriate sections of Tables \ref{tab:distances} and \ref{tab:comparisons} and other results measured, we see a substantially improved model. As with Model 1, we have results that appear graphically similar across the entirety or keys sections of the distribution for active links, active nodes, global clustering coefficient, on-durations and off-durations, whilst our Kolmogorov-Smirnov distances for these indicate that there are still improvements to the fits to be made here. For our triangle count, we are seeing reasonable fits graphically and are accepting a higher number of our statistical comparisons.
Again, for nodes per component, links per component and the component count, we observe acceptable fits graphically (partially presented in Figure \ref{fig:model2a}) and are indeed accepting a small number of these fits when calculating our statistical distances - overall a slightly higher number than in Model 1, but with only small variations in each one. For node activity potential, we have a very good fit, both graphically and when considering Kolmogorov-Smirnov distances.
As can be seen when we compare Figure \ref{subfig:echoes_orig} and Figure \ref{subfig:echoes_m2a}, we are also producing an acceptable link selection spread (defined in section {sec:modelanalysis}), which reflects the varying levels of friendships observed in the real world data.

However, this model is insufficient to capture the related component structure - with our generated data still having too many linear components in comparison to triangles. Although attempting to resolve this will increase our dependence on the data, it is believed to be significant enough to warrant this.

\subsection{Model 2b}\label{ssec:model2b}
\begin{figure}
\centering
\includegraphics[width=0.85\columnwidth]{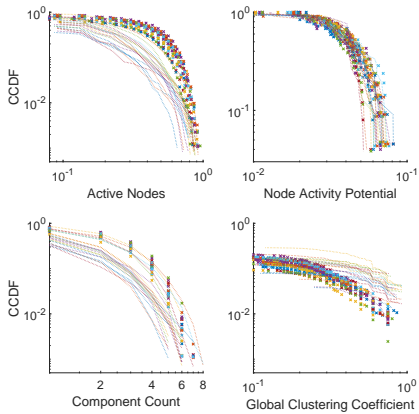}
\caption{Selected Results for Model 2b. Simulated data is represented by crosses whereas observed data is represented by dotted lines. Each colour represents a different simulation or data set. See subsection \ref{ssec:model2b} for full explanation.}
\label{fig:model2b}
\end{figure}
In Figure \ref{fig:model2b}, the relevant sections of our tables and other results measured, we see similar results to Model 2a. Again, we have fits that have various levels of visual similarity to the observed data for active links, active nodes, on-durations and off-durations, whilst our Kolmogorov-Smirnov distances as reported in Tables \ref{tab:distances} and \ref{tab:comparisons}, for these indicate that there are issues with these. With our global clustering coefficient have reasonable fits graphically, but similar to Model 2a are still having issues with Kolmogorov-Smirnov acceptances.
Again, for nodes per component, links per component and the component count, we observe acceptable fits graphically (partially presented in Figure \ref{fig:model2b}) and note in Table \ref{tab:comparisons} a slight increase or similar levels in count of acceptances. We have a similar result for the node activity potential, with a very good graphic fit and a very high number of Kolmogorov-Smirnov acceptances.
For the triangle count in the network, we observe good fits graphically and in terms of our statistical tests, with a substanial improvement over the results obtained in Model 2a.
We also observe varying levels of popularity in the links, reflecting the various levels of friendships that can be seen in the original data - as can be seen in comparing behaviours in Figure \ref{subfig:echoes_orig} and Figure \ref{subfig:echoes_m2a}.

\subsection{Model 2c}\label{ssec:model2c}
\begin{figure}
\centering
\includegraphics[width=0.85\columnwidth]{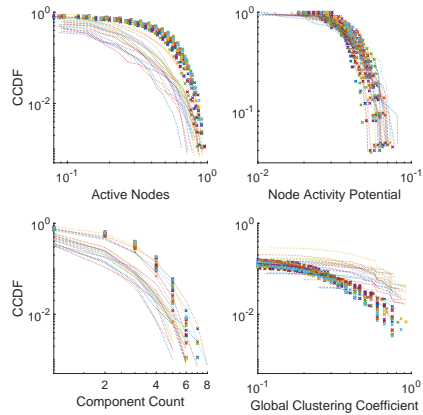}
\caption{Selected Results for Model 2c. Simulated data is represented by crosses whereas observed data is represented by dotted lines. Each colour represents a different simulation or data set. See subsection \ref{ssec:model2c} for full explanation.}
\label{fig:model2c}
\end{figure}
In most metrics, this model performs similarly to Model 2b, with little to no difference in all of our examined metrics. Whilst, as illustrated in Table \ref{tab:comparisons}, some see a slight drop in the number of acceptances of the null hypothesis for the two-sample Kolmogorov-Smirnov test, others see a slight increase and overall we see a very marginal increase in the total count. Overall behaviours and link selection weighting reflect the observed data with a reasonable degree of accuracy as can be seen in Figure \ref{fig:model2c} and a comparison between Figures \ref{subfig:echoes_orig} and \ref{subfig:echoes_m2a}.

\section{Model Comparison}\label{sec:modelcomp}
\begin{table*}
\centering
\begin{tabular}{l|c|c|c|c|} \cline{2-5}
& \textbf{Model 1} & \textbf{Model 2a} & \textbf{Model 2b} & \textbf{Model 2c}\\ \hline
\multicolumn{1}{|l|}{\textbf{Active Links}} & $1$ & $0$ & $1$ & $0$\\ \hline
\multicolumn{1}{|l|}{\textbf{Active Nodes}} & $0$ & $0$ & $0$ & $0$\\ \hline
\multicolumn{1}{|l|}{\textbf{Node Activity Potential}} & $954$ & $1000$ & $1000$ & $999$\\ \hline
\multicolumn{1}{|l|}{\textbf{Global Clustering Coefficient}} & $0$ & $0$ & $0$ & $0$\\ \hline
\multicolumn{1}{|l|}{\textbf{Interaction Time}} & $6$ & $0$ & $0$ & $0$\\ \hline
\multicolumn{1}{|l|}{\textbf{Time Between Contacts}} & $24$ & $2$ & $2$ & $2$\\ \hline
\multicolumn{1}{|l|}{\textbf{Component Count}} & $151$ & $264$ & $240$ & $306$\\ \hline
\multicolumn{1}{|l|}{\textbf{Links per Component}} & $62$ & $35$ & $56$ & $42$\\ \hline
\multicolumn{1}{|l|}{\textbf{Nodes per Component}} & $33$ & $68$ & $79$ & $63$\\ \hline
\multicolumn{1}{|l|}{\textbf{Triangle Count}} & $152$ & $503$ & $637$ & $637$\\ \hline
\end{tabular}

\caption{Acceptances of $\mathcal{H}_{0}$ (see equation \ref{eq:h0comparisons}) at $5\%$ Level (max: 1000) (see section \ref{sec:modelanalysis} for full explanation)}
\label{tab:comparisons}
\end{table*}
Overall, there is a considerable improvement across most metrics between Model 1 and Model 2a. This can be seen empirically when we examine the statistical distances between the observed data and our generated simulations and the count of $5\%$ acceptances (as illustrated in Tables \ref{tab:distances} and \ref{tab:comparisons}).  Significant improvements are made to the node activity potential and triangle count, including a noticeable graphical improvement to the global clustering coefficient, as can been seen in Figures \ref{fig:model1} and \ref{fig:model2a}. Whilst modifications could be made to Model 1 to improve its accuracy in some of these areas (such as including the link selection preference matrix), due to its improved performance with similar levels of dependence on the data, the second model will be the basis for all future work. We also notice a substantial drop in link selection spread (defined in section {sec:modelanalysis}) as we move between these models, with Model 2a reflecting real world behaviours much closer in our observations, as displayed in Figure \ref{fig:echoes}.

Between Model 2a and Model 2b, many metrics remain similar, although as expected from our modifications to the algorithm, we do notice a considerable improvement to the triangle count, illustrated in both Table \ref{tab:comparisons} and when observing the decrease in the maximum and mean statistical distance for this metric in Table \ref{tab:distances}. However, one of the larger problems with Model 2b is that the link selection preference matrix depends heavily on the original data, and we note that we could reduce this data draw considerably by generating this matrix rather than extracting it directly from the data. Model 2c attempts to do this, and can be considered successful as we can observe in Tables \ref{tab:distances} and \ref{tab:comparisons}, although a deeper examination of the temporal and network properties indicates that further improvements are still to be made.

\section{Model Validation}\label{sec:validation}
\begin{table*}
\centering
\begin{tabular}{l|c|c|c|} \cline{2-4}
& \textbf{Model 1} & \textbf{Model 2a} & \textbf{Model 2b}\\ \hline
\multicolumn{1}{|l|}{\textbf{Active Links}} & $0$ & $2$ & $0$\\ \hline
\multicolumn{1}{|l|}{\textbf{Active Nodes}} & $56$ & $389$ & $362$\\ \hline
\multicolumn{1}{|l|}{\textbf{Node Activity Potential}} & $881$ & $1000$ & $1000$\\ \hline
\multicolumn{1}{|l|}{\textbf{Global Clustering Coefficient}} & $0$ & $0$ & $0$\\ \hline
\multicolumn{1}{|l|}{\textbf{Time Between Contacts}} & $-$ & $2$ & $2$\\ \hline
\multicolumn{1}{|l|}{\textbf{Component Count}} & $0$ & $211$ & $364$\\ \hline
\multicolumn{1}{|l|}{\textbf{Links per Component}} & $0$ & $43$ & $45$\\ \hline
\multicolumn{1}{|l|}{\textbf{Nodes per Component}} & $0$ & $42$ & $48$\\ \hline
\multicolumn{1}{|l|}{\textbf{Triangle Count}} & $0$ & $367$ & $771$\\ \hline
\end{tabular}

\caption{Validation Acceptances of $\mathcal{H}_{0}$ (see equation \ref{eq:h0validation}) at $5\%$ Level (max: 1000) (see subsection \ref{sec:validation} for full explanation)}
\label{tab:validation}
\end{table*}
As indicated in Table \ref{tab:comparesamples}, our approach to using the same distributions through all of our primary school models is not ideal. Therefore, we shall examine our methods in such a way to examine if the dynamics used in our models is a valid choice. To do this we shall draw temporal data directly from the appropriate eCCDFs - for Model 1, these are the on-, off- and activation times, whilst for Model 2, these are the on-times and interevent times. We shall then compare the data generated using this method to the real world sample that we draw the eCCDFs from - if we have a low statistical distance between these, we can conclude that our model dynamics have validity and that any issues identified in the examination above can be significantly addressed through parameter improvements and refinements to the choice of distributions for our random values.

In Table \ref{tab:validation}, we present the results of our validation. We take each of our 20 original data samples and input the appropriate eCCDFs in the place of the random generation outlined in Methods 1, 2a and 2b as described in section \ref{sec:models}. We do not analyse Method 2c using this method of validation as if we were to draw the link preferential matrix in this method from the data, this would be functionally identical to Model 2b.

We then generate 50 samples for each and compare them to the original data (for a total of 1000 comparisons for each metric and model). Please node that interaction times for all models (and the time between contacts for Model 1) have been excluded from this table as they are being controlled directly from the data and thus, a validation using this metric would serve no purpose. In this table our $\mathcal{H}_{0}$ is given as
\begin{equation}\label{eq:h0validation}
\begin{split}
\mathcal{H}_{0}:  &\text{ The chosen observed and validation}\\
									&\text{ samples come from a common}\\
									&\text{ distribution.}
\end{split}
\end{equation}

Using this data, we can clearly see that our variations of Model 2 have considerably improved dynamics over Model 1, although we can still see that there are still improvements to be made. When we compare Tables \ref{tab:comparisons} and \ref{tab:validation}, we observe whilst choosing the `right' time structures does lead to some improvements - most notably in terms of active nodes - it is not enough to ensure a fit across all chosen metrics, and therefore that changes to our overall dynamics should be considered. From our examination of these results, we conclude that efforts should be made to improve link dynamics and hypothesise that by modifying our code to change the number of links generated in the network should improve our dynamics - especially for active nodes, although we would expect to also see improvements in our global clustering coefficient, component features and active nodes. This should also improve our time between contacts as changing the number of link activations will have a direct impact on this metric.

However, despite these small improvements still to be made to our model, we conclude that our Model 2b (and therefore 2c) have justifiable dynamics and that an improvement to the random generations will lead to an improved model overall.

\section{Conclusions}
We have developed two forms of model for the social interactions observed in the original data collected by the SocioPatterns collaboration \cite{Gemmetto2014,10.1371/journal.pone.0023176,10.1371/journal.pone.0136497}. In terms of statistical distance, both of these exhibit varying degrees of matching with the original data - the second of our models out-performing the first in almost all of our chosen metrics. We have added refinements to this, improving upon this matching, whilst continuing to minimise the amount of dependence on the underlying data.
We also have run a form of model validation and can certainly acknowledge that our model dynamics have a notable degree of validity in a number of key metrics when compared to the real word data - whilst this indicates that we do have additional improvements to the mechanisms in our models to perform, we believe that our current models are a promising step in a strong direct. 
We also acknowledge that further refinement for the parameters and distributions used may lead to improved matching, although we believe the models presented here provide a solid foundation from which to proceed.

\section{Future Work}
Improvements to the method for generating our matrix in Model 2c will have to be undertaken before this algorithm is finalised. Additionally, further parameter and distribution refinement in our method will also be explored, including potentially moving from a Log-Normal distribution for the interevent times to a more complicated method in order to improve the matching between the generated time between contacts and that in the real world data. We will also attempt to make modifications as proposed in our model validation in section \ref{sec:validation}, although these improvements are only hypothesised to improve model dynamics. Once we have completed our model for primary school data, we shall move to the high school data by using the same method and adjusting parameters.

We will also carry out a deeper theoretical analysis of our model and examine any interesting patterns or behaviours within it, looking at long-term behaviours and, through simulation, the potential existence of any absorbing states. Additionally, once we have a finalised model and thus a statistically rigorous understanding of the distributions behind the observed behaviours, we can propose theoretical reasoning for these choices by examining the significance and underlying mechanisms of such distributions.

Eventually, we aim to place a network-driven epidemic model on our time-varying network and examine properties of disease spread and potential predictive power, comparing both to existing models and real data.

\section*{Notes}\label{sec:notes}
Supplemental materials can be found at the following link: {\tt https://drive.google.com/drive/folders/}\\
\noindent {\tt 1nLpdt91XUNElF1es2x3sm6qemqkQGrGf?usp=sharing}

\section*{Acknowledgements}
We acknowledge useful discussion with J\'{a}nos Kert\'{e}sz at the 13th Econophysics Colloquium \& 9th Polish Symposium on Physics in Economy and Social Sciences, Warsaw, July 2017. Additionally, we would like to thank and anonymous referee for a thorough reading of this article and for the useful suggestions that improved the clarity and presentation. This research has been partially funded by an EPSRC DTP grant.
\bibliography{BibTeX1}
\end{document}